\documentclass[11pt]{article} 
\usepackage{mystyle-new}
\usepackage{authblk}
\usepackage[T1]{fontenc}
\usepackage{epsfig,amsmath} 
\usepackage{hepnames,hepunits}
\usepackage{hyperref}
\usepackage{color}
\usepackage{graphicx}
\definecolor{red}{rgb}{1,0,0}
\def\lesssim{\ \hbox{\raise 2pt \hbox{$<$} \kern -13pt
                     \lower 3pt \hbox{$\sim$}}\ }
\def\greatersim{\ \hbox{\raise 2pt \hbox{$>$} \kern -13pt
                     \lower 3pt \hbox{$\sim$}}\ }

\def\lsim{\mathrel{\rlap{\lower4pt\hbox{\hskip1pt$\sim$}}
    \raise1pt\hbox{$<$}}}                
\def\gsim{\mathrel{\rlap{\lower4pt\hbox{\hskip1pt$\sim$}}
    \raise1pt\hbox{$>$}}}                

\def\cascade{{\sc Cascade}}
\def\pythia{{\sc Pythia}}
\def\herwig{{\sc Herwig}}

\def\mcatnlo{{MCatNLO}}

\input epsf.tex
\def\desepsf(#1 width #2){\epsfxsize=#2 \epsfbox{#1}}
\def\kt{\ensuremath{k_{\rm T}}}

\def\pt{\ensuremath{p_{\rm T}}}
\def\ptjj{\ensuremath{p_{{\rm T},12}}}

\newcommand{\alphas}{\ensuremath{\alpha_\mathrm{s}}}

\newcommand{\asmz}{\ensuremath{\alphas(m_{\PZ})}\xspace}

\newcommand{\PBM}{PB}
\newcommand{\TMDlib}{{\sc TMDlib}}

\newcommand{\dphi}{{\ensuremath{\Delta\phi_{12}}}}
\newcommand{\MCatNLO}{{\sc MadGraph5\_aMC@NLO}}
\newcommand{\ptmax}{\ensuremath{\pt^{\text{leading}}}\xspace}

\newenvironment{tolerant}[1]{\par\tolerance=#1\relax}{ \par }
\usepackage{amsmath,bm}
\usepackage{lineno}

\usepackage{cite,./mcite}
\usepackage{tikz}
\usepackage[symbol]{footmisc}

\providecommand{\DOI}[1]{\href{http://dx.doi.org/#1}}

\begin{document}

\title{
Azimuthal correlations of high transverse momentum jets at next-to-leading order in the parton branching method }
\author[1]{M.~I.~Abdulhamid}
\affil[1]{\small Department of Physics, Tanta University, Tanta, Egypt}
\author[2]{M. A. Al-Mashad}
\affil[2]{Center for High Energy Physics (CHEP-FU), Faculty of Science, Fayoum University, Egypt}
\author[3]{A.~Bermudez~Martinez}
\affil[3]{Deutsches Elektronen-Synchrotron DESY, Germany}
\author[4]{G.~ Bonomelli}
\affil[4]{Università degli Studi Milano-Bicocca, Italy}
\author[5,6]{I.~ Bubanja}
\affil[5]{Interuniversity Institute for High Energies (IIHE), Universit\'e libre de Bruxelles, Belgium}
\affil[6]{Faculty of Natural sciences and Mathematics, University of Montenegro, Podgorica, Montenegro}
\author[7]{N.~Crnković}
\affil[7]{Ruđer Bošković Institute, Zagreb, Croatia}
\author[3]{F.~Colombina}
\author[8]{B.~D’Anzi}
\affil[8]{National Institute for Nuclear Physics INFN and University of Bari, Italy}
\author[9]{S.~Cerci}
\affil[9]{Adiyaman University, Adiyaman, Turkey}
\author[10]{M.~Davydov}
\affil[10]{Moscow State University, Russia}
\author[3]{L.I.~Estevez~Banos}
\author[4]{N.~ Forzano}
\author[11,12]{F.~Hautmann}
\affil[11]{Elementary Particle Physics, University of Antwerp, Belgium}
\affil[12]{University of Oxford, UK}
\author[3]{H.~Jung}
\author[12]{S.~Kim}
\author[13]{A.~León~Quirós}
\affil[13]{Universidad Autónoma de Madrid, Madrid, Spain}
\author[14]{D.~E.~ Martins}
\affil[14]{Universidade Federal de Pelotas, Brazil}
\author[3]{M.~Mendizabal}
\author[15]{K. Moral Figueroa}
\affil[15]{University of Edinburgh, UK}
\author[16] {S.~Prestel}
\affil[16]{Department of Astronomy and Theoretical Physics, Lund University, Sweden} 
\author[3]{S.~Taheri~Monfared}
\author[17]{C.~S\"usl\"u}
\affil[17]{Department of Physics, Ihsan Dogramaci Bilkent University, Ankara, Turkey}
\author[9]{D. Sunar Cerci}
\author[11]{A.M.~van~Kampen}
\author[11]{P.\@ Van Mechelen}
\author[18]{A.~Verbytskyi}
\affil[18]{Max-Planck Institut f\"ur Physik, Munich, Germany}
\author[3,19]{Q.~Wang}
\affil[19]{School of Physics, Peking University}
\author[3,19]{H.~Yang}
\author[20]{Y.~Zhou}
\affil[20]{University of Cambridge, UK}

\begin{titlepage} 
\maketitle
\vspace*{-21cm}
\begin{flushright}
DESY-21-219 \\
LU-TP 21-53 \\ 
\today
\end{flushright}
\vspace*{+17cm}
\end{titlepage}

\begin{abstract}
The azimuthal correlation, \dphi , of high transverse momentum jets in $\Pp\Pp$ collisions at $\sqrt{s}=13$ \TeV\ is studied by applying  \PBM -TMD distributions to NLO calculations via \mcatnlo\  together with the \PBM -TMD parton shower. 
A very good description of the cross section as a function of \dphi\ is observed. In the back-to-back region of $\dphi \to \pi$, a very good agreement is observed with the \PBM -TMD Set~2 distributions while significant deviations are obtained with the  \PBM -TMD Set~1 distributions. Set~1 uses the evolution scale while Set~2 uses transverse momentum as an argument in \alphas, and the above observation therefore confirms the importance of an appropriate soft-gluon coupling in angular ordered parton evolution.
The total uncertainties of the predictions are dominated by the scale uncertainties of the matrix element, while the uncertainties coming from the \PBM -TMDs and the corresponding \PBM -TMD shower are very small.  
The  \dphi\ measurements are also compared with predictions using  \mcatnlo\  together \pythia 8, illustrating the importance of details of the parton shower evolution.
\end{abstract}

\section{Introduction} 
\label{Intro}
The description of the cross section of high \pt\ jets in proton-proton ($\Pp\Pp$) collisions is one of the most important tests of predictions obtained in Quantum Chromodynamics (QCD), and much progress has been achieved in the description of inclusive jets \cite{ATLAS:2013pbc,CMS:2015jdl,CMS:2012ftr,CMS:2014nvq,ATLAS:2017kux,Khachatryan:2016mlc,Aaboud:2017wsi,Khachatryan:2016wdh,Sirunyan:2020uoj,Acharya:2019jyg,Aad:2011fc,ATLAS:2010dmq} 
 by applying next-to-leading (NLO) ~\cite{Alioli:2010xa,Nagy:2001fj,Giele:1994gf,Ellis:1992en} and next-to-next-to-leading-order (NNLO) calculations~\cite{Czakon:2019tmo,Gehrmann-DeRidder:2019ibf,Currie:2016bfm,Currie:2017ctp}.
 In multijet production, the azimuthal angle \dphi\  between the two highest transverse momentum  \pt -jets is an inclusive measurement of additional jet radiation. At leading order (LO) in strong coupling \alphas , where only two jets are present, the jets are produced back-to-back, with $\dphi = \pi$, while a deviation from this back-to-back configuration indicates the presence of additional jets, and only higher-order calculations can describe the observations. The azimuthal correlation between two jets has been measured in $\Pp\Pap$ collisions at a center-of-mass energy of $\sqrt{s}=1.96$~\TeV\ by the D0 collaboration~\cite{Abazov:2004hm,D0:2012dqa} and in $\Pp\Pp$ collisions by the ATLAS Collaboration at $\sqrt{s}=7$~\TeV~\cite{daCosta:2011ni} and by the CMS Collaboration at $\sqrt{s}=7$, $8$, and $13$~\TeV~\cite{Khachatryan:2011zj,Khachatryan:2016hkr,Sirunyan:2017jnl,Sirunyan:2019rpc}. When measurements of azimuthal correlations of dijets are compared with LO or NLO computations supplemented 
by parton showers, deviations of 50\% are observed in the medium \dphi\ region even at NLO (see e.g. \cite{Khachatryan:2016hkr,Sirunyan:2017jnl}), which requires a more detailed understanding.
 In  the  $\dphi \to \pi$ region, deviations of  up to 10 \%  are observed~\cite{Sirunyan:2019rpc}, significantly larger than the experimental uncertainties. 

Since initial state parton radiation moves the jets away from the $\dphi = \pi$ region, it is appropriate to investigate the implications of  transverse momentum dependent parton densities (TMDs \cite{Angeles-Martinez:2015sea}) in the description of the \dphi\ measurements. 
Kinematic effects of the initial-state transverse momenta in the interpretation of jet measurements were  pointed out in \cite{Dooling:2012uw,Hautmann:2012dw}. 
The region $\dphi \to \pi$ is especially sensitive to soft multi-gluon emissions, for which QCD resummation is needed. Calculations at leading-logarithm have been obtained in Ref.~\cite{Banfi:2008qs}. 
A calculation based on  TMD distributions is found in Ref.~\cite{Sun:2014gfa,Sun:2015doa} and further investigated in \cite{Hatta:2021jcd,Hatta:2020bgy}. 
However, in the  region $\dphi \to \pi$ soft-gluon effects are expected which lead to so-called  {\it factorization - breaking}~\cite{Collins:2007nk,Vogelsang:2007jk,Rogers:2010dm}.  An indirect strategy to explore the potential  impact of these effects is to compare calculations which assume factorization with high-precision measurements.
 
The Parton Branching (\PBM ) - method~\cite{Hautmann:2017xtx,Hautmann:2017fcj} allows one to determine TMD parton distributions. With these \PBM -TMD distributions a very good description of the Drell-Yan process~\cite{Drell:1970wh} is achieved  at the LHC~\cite{Martinez:2019mwt}  as well as at lower energies~\cite{Martinez:2020fzs}. Drell-Yan lepton pair production in association with jet final states is also well described  using the TMD jet merging~\cite{Martinez:2021chk}.  
In Ref.~\cite{Baranov:2021yut} it is shown that $\PZ + \Pqb$ production is also well described. TMD parton distributions have been used together with off-shell matrix elements at the lowest order in Refs.~\cite{Hautmann:2008vd,Dooling:2014kia,Bury:2017jxo} showing a reasonably good description of the measurements.

In this article we investigate in detail high-\pt\  dijet production by applying the \PBM\ formulation of TMD evolution together with NLO calculations of the hard scattering process in the  \MCatNLO\ \cite{Alwall:2014hca} framework. We first give a very brief recap of the \PBM\ distributions in Sec.\@~\ref{PBTMD}. In Sec.\@~\ref{sec:aMCatNLO} we describe how TMDs and TMD parton showers are included in the Monte Carlo generator \cascade3~\cite{Baranov:2021uol}. We discuss predictions obtained by applying fixed-order NLO perturbative calculations and study the region where soft gluon resummation becomes important. We show predictions using \PBM\ - TMDs together with TMD parton shower in Sec.\@~\ref{sec:correlations}. We compare these predictions with the one using the \pythia8\ parton shower.
 We finally give conclusions in Sec.\@~\ref{sec:concl}.

\section{PB TMDs}
\label{PBTMD}
The PB method~\cite{Hautmann:2017fcj,Hautmann:2017xtx}  provides a solution of evolution equations for collinear and TMD parton distributions. The equations are solved by applying the concept of resolvable and non-resolvable branchings with Sudakov form factors providing the probability to evolve from one scale to another without resolvable branching.
The method is described in Refs.~\cite{Martinez:2018jxt,Hautmann:2017fcj}. 

\begin{figure}[h!tb]
\begin{center} 
\includegraphics[width=0.48\textwidth]{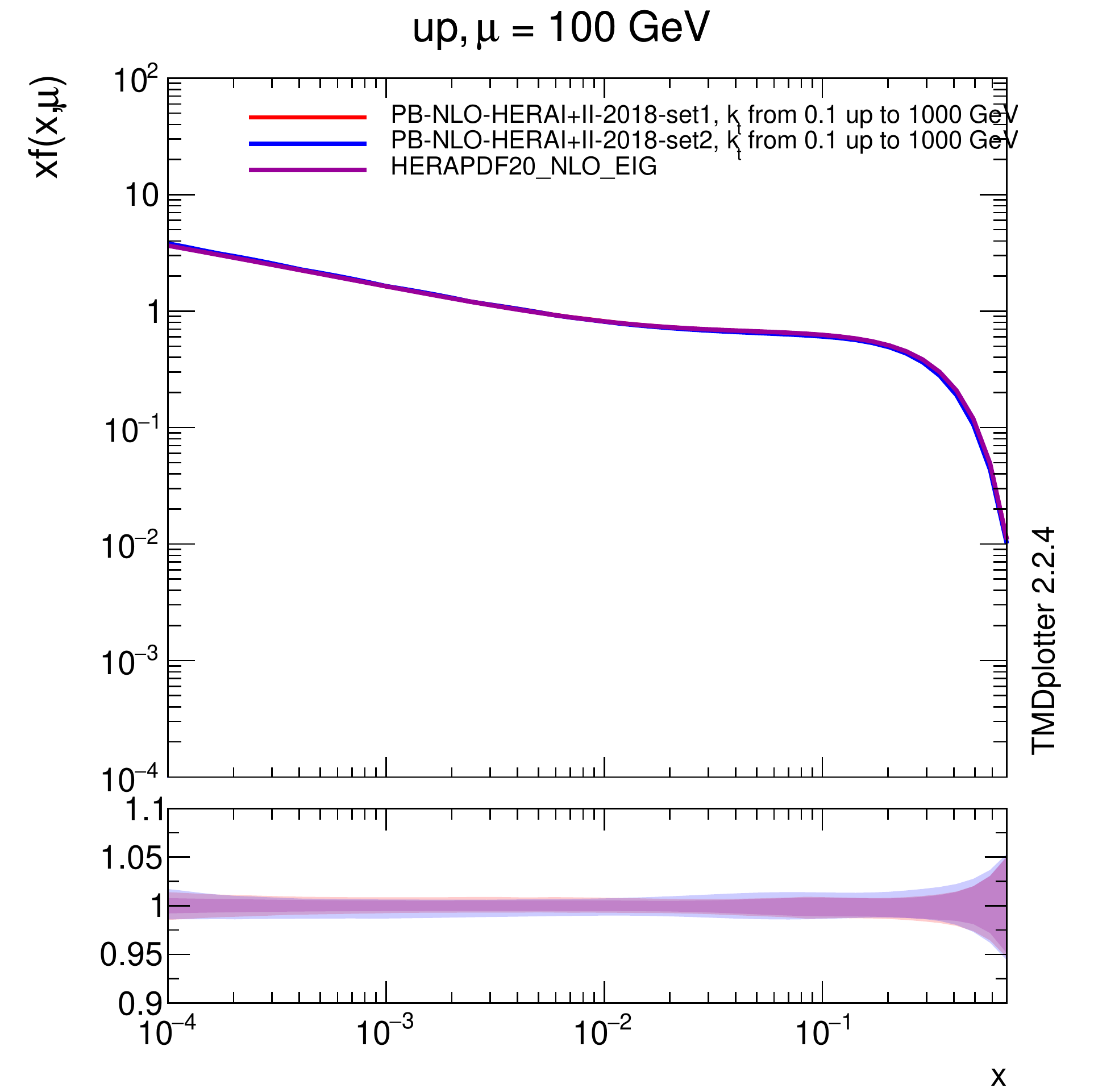} 
\includegraphics[width=0.48\textwidth]{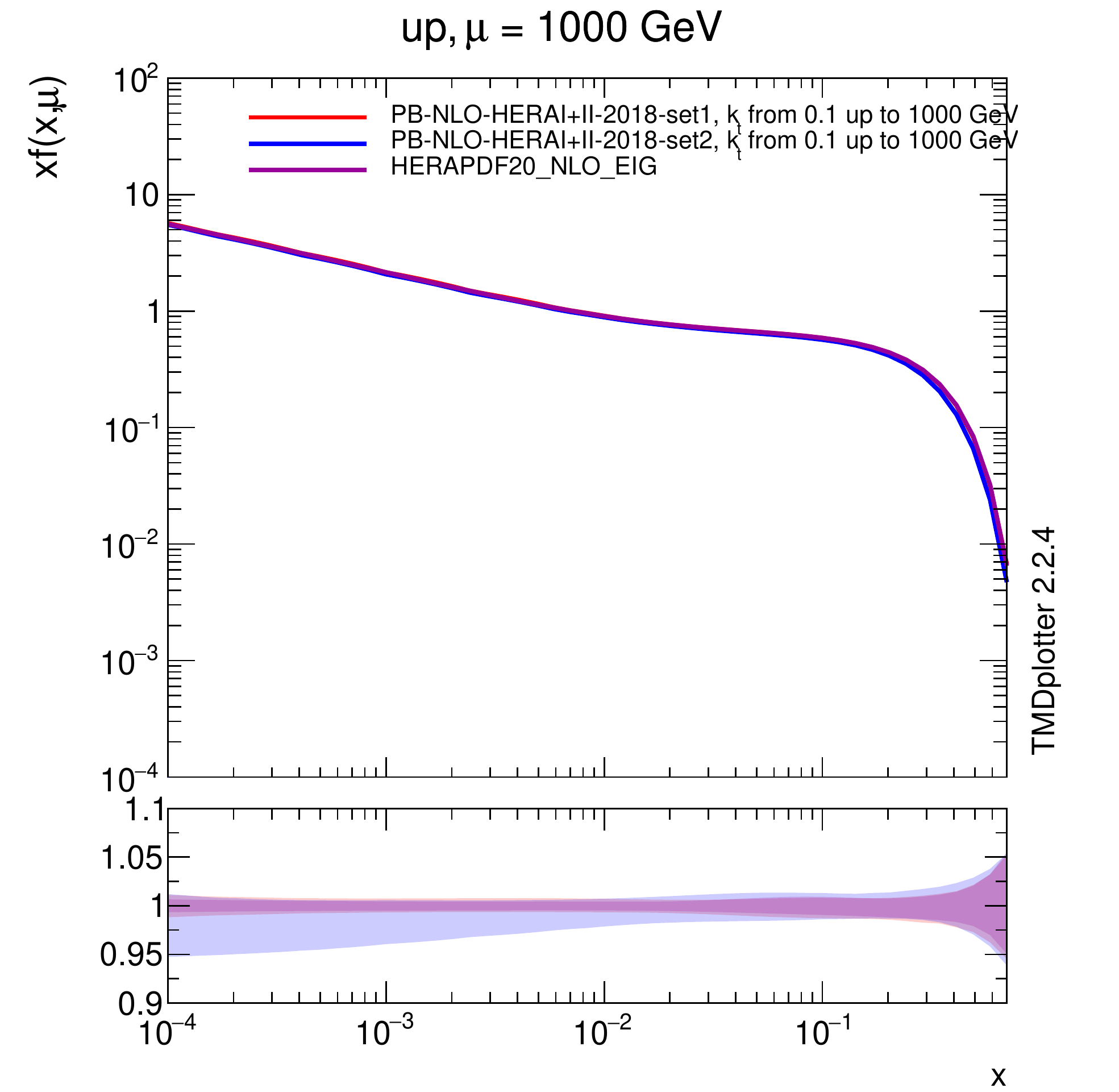} 
  \caption{\small Collinear parton density distributions for up quarks  (PB-NLO-2018-Set 1, PB-NLO-2018-Set 2 and HERAPDF2.0) as a function of $x$ at   $\mu = 100 $ and $1000$ \GeV. In the lower panel the uncertainties are shown.
  }
\label{TMD_pdfs1}
\end{center}
\end{figure} 

\begin{tolerant}{500}
For the numerical calculations we use the NLO parton distribution sets, PB-NLO-2018-Set 1 and PB-NLO-2018-Set 2, as obtained in Ref.~\cite{Martinez:2018jxt} from a fit to inclusive deep inelastic scattering precision measurements at HERA~\cite{Abramowicz:2015mha}. Both the collinear and TMD distributions are available in \TMDlib ~\cite{Abdulov:2021ivr,Hautmann:2014kza}, including uncertainty bands. PB-NLO-2018-Set 1 corresponds at collinear level to HERAPDF 2.0 NLO~\cite{Abramowicz:2015mha},  while PB-NLO-2018-Set 2 uses transverse momentum (instead of the evolution scale in Set1) for the scale in the running coupling $\alphas$ which corresponds to the angular ordering of soft gluon emissions in the initial-state parton evolution~\cite{Bassetto:1983mvz,Dokshitzer:1987nm,Catani:1990rr,Hautmann:2019biw}.
\end{tolerant}
\begin{figure}[h!tb]
\begin{center} 
\includegraphics[width=0.45\textwidth]{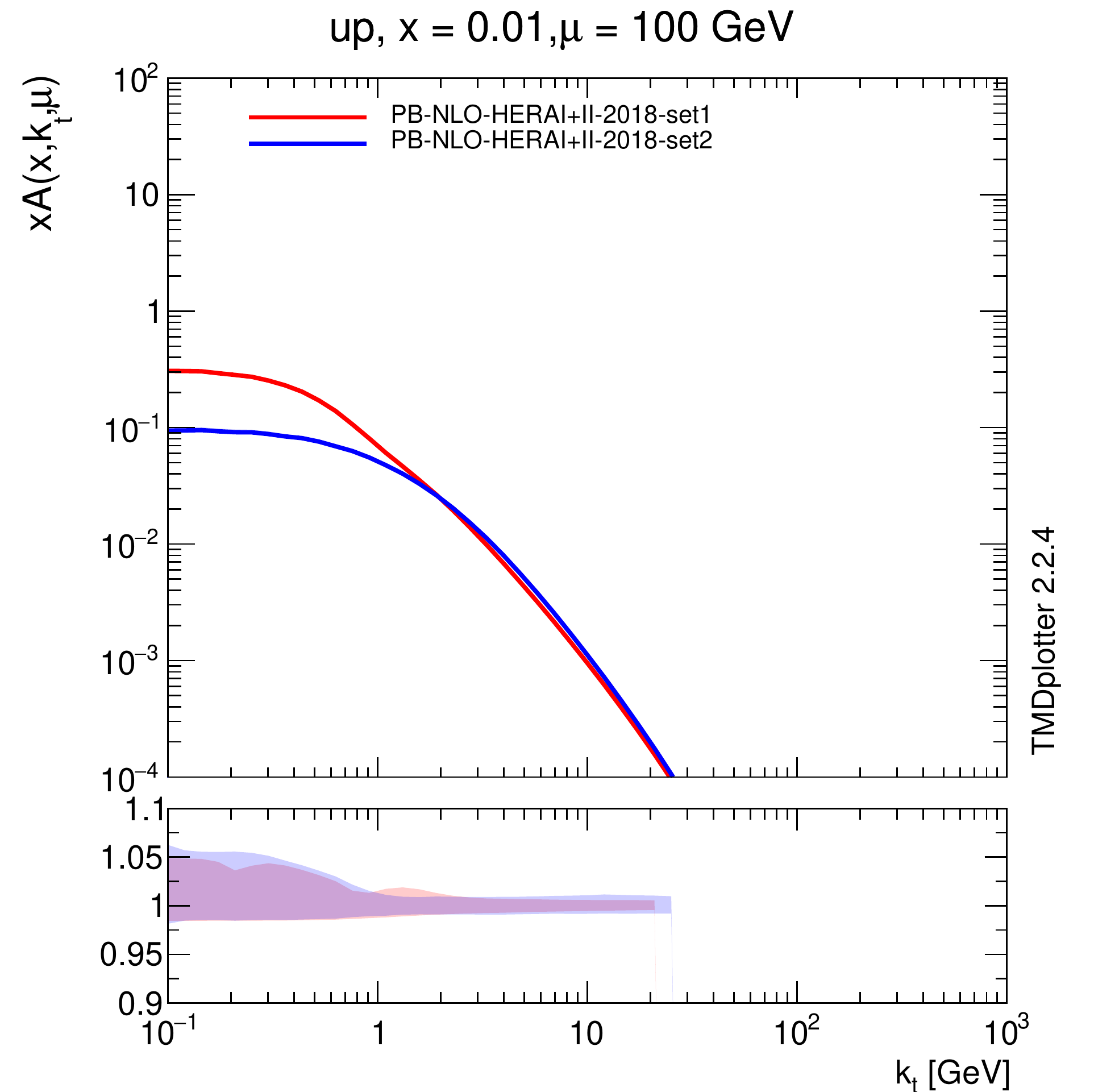} 
\includegraphics[width=0.45\textwidth]{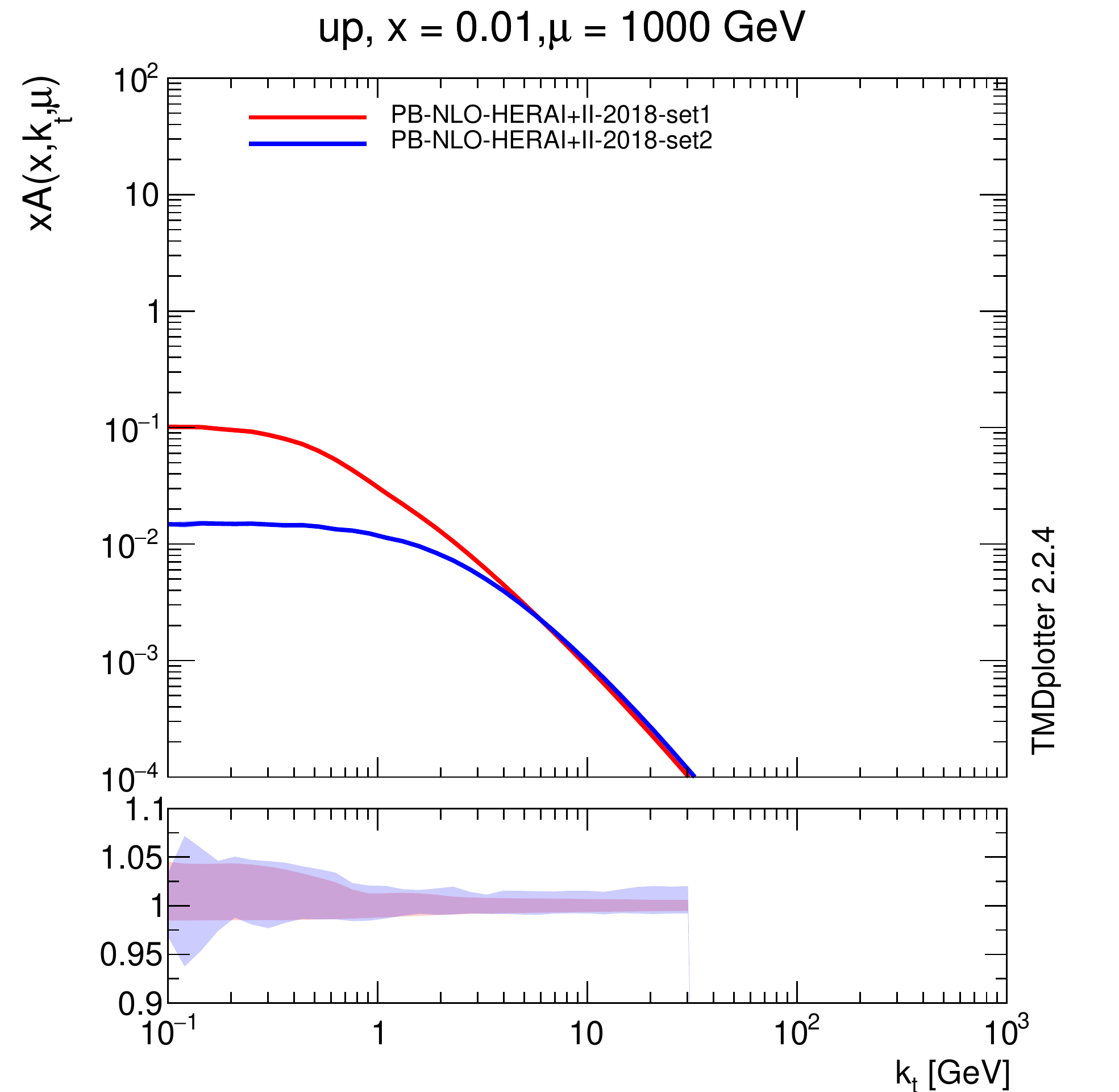} 
  \caption{\small TMD parton density distributions for up quarks  (PB-NLO-2018-Set 1 and PB-NLO-2018-Set 2) as a function of $\kt$ at $\mu=100$ and $1000$  \GeV\  and $x=0.01$.
  In the lower panels show the full uncertainty of the TMDs, as obtained from the fits \protect\cite{Martinez:2018jxt}.
  }
\label{TMD_pdfs2}
\end{center}
\end{figure}

In Fig.~\ref{TMD_pdfs1} the distributions of the collinear densities from  Set~1 and Set~2 are shown for up-quarks at  evolution scales of $\mu=100$ and $1000$  \GeV , typical for multi-jet production described below.
The collinear densities are also available in a format compatible with LHAPDF~\cite{Buckley:2014ana},  and can be used in calculations of physical processes at NLO.  
In Fig.~\ref{TMD_pdfs2}   we show the TMD distributions for up-quarks at $x=0.01$ and $\mu=10$ and $100$ \GeV . The differences between Set~1 and Set~2 are clearly visible in the small \kt -region.

The uncertainties of the TMD distributions include both experimental and model uncertainties, as determined in Ref.~\cite{Martinez:2018jxt}. In general, it is observed that those uncertainties are small; for $\kt > 1$ \GeV\ they are of the order of 2--3 \%.

\section{Multijet production}
\label{sec:aMCatNLO}
\begin{tolerant}{7000}
The predictions for multijet production at NLO are obtained using the \MCatNLO \cite{Alwall:2014hca} framework. We used \MCatNLO\ in two different modes: one is the fixed NLO mode, in which only partonic events are produced, without parton shower and hadronization, and the other one is the real MC@NLO mode, in which 
infrared subtraction terms are included to avoid double counting of parton emissions between matrix-element and parton-shower calculations, so that events need to be supplemented with a parton shower (or with PB TMD evolution) in order to produce a physical cross section.
 \end{tolerant}

\begin{tolerant}{500}
Fixed NLO dijet production is calculated within the \MCatNLO\ framework.   Technically, in the fixed NLO mode, \MCatNLO\  (version 2.9.3)  produces event files with the partonic configuration in LHE format~\cite{Alwall:2006yp}  which can be processed through \cascade 3~\cite{Baranov:2021uol}  combining events and counter events (due to infrared subtraction) so that they are treated as one  event  for a proper calculation of statistical uncertainties. In the fixed NLO mode, the  \MCatNLO\  event record is kept without any modification. Processing through \cascade 3 has a significant advantage that a fixed NLO calculation can be obtained making use of all the analyses coded in  Rivet~\cite{Buckley:2010ar}.
\end{tolerant}

In the MC@NLO mode, subtraction terms are included which depend on the parton shower used. For the \PBM -TMDs and the \PBM -TMD parton shower we use \herwig6  \cite{Corcella:2002jc,Marchesini:1991ch}  subtraction terms, as already applied in \PZ and Drell-Yan analyses~\cite{Martinez:2020fzs,Martinez:2019mwt}, motivated by the angular ordering in the PB evolution. \MCatNLO\  (version 2.6.4, hereafter  labeled \mcatnlo )  \cite{Alwall:2014hca} together with the NLO \PBM\ parton distributions with $\alphas(M_{\PZ } )= 0.118$ is used for NLO calculation of dijet production. The matching scale $\mu_{m}$, which limits the contribution from \PBM -TMDs and TMD showers ($\mu_{m}=\verb+SCALUP+$ included  in the LHE file), guarantees that the overlap  with real emissions from the matrix element is minimized according to the subtraction of counterterms in the MC@NLO method.
The factorization and renormalization scale in \mcatnlo\ is set to  $\mu_{R,F}=\frac{1}{2} \sum_i \sqrt{m^2_i +p^2 _{t,i}}$, where the index $i$ runs over all particles in the matrix element final state. This scale is also used in the \PBM -TMD parton distribution ${\cal A}(x,\kt,\mu)$.

In \cascade 3, as described in detail in Ref.~\cite{Baranov:2021uol}, the transverse momentum of the initial state partons is calculated according to the distribution of \kt\ provided by the  \PBM -TMD ${\cal A}(x,\kt,\mu)$ at given longitudinal-momentum fraction $x$ and evolution scale $\mu$. This transverse momentum is used for the initial state partons provided by \mcatnlo , and their longitudinal momentum is adjusted  such that the mass and the rapidity of the dijet system is conserved, similar to what has been done in the Drell-Yan case~\cite{Martinez:2020fzs}. The initial state TMD parton shower is included in a backward evolution scheme, respecting all parameters and constraints from the \PBM -TMD. The kinematics of the hard process are not changed by the shower, after the \kt\ from the TMD is included. The final state parton shower is obtained with the corresponding method implemented in \pythia 6~\cite{Sjostrand:2006za}, by vetoing emissions which do not satisfy angular ordering (\verb+MSTJ(42)=2+).

\begin{figure}[h!tb]
\begin{center} 
\includegraphics[width=0.45\textwidth]{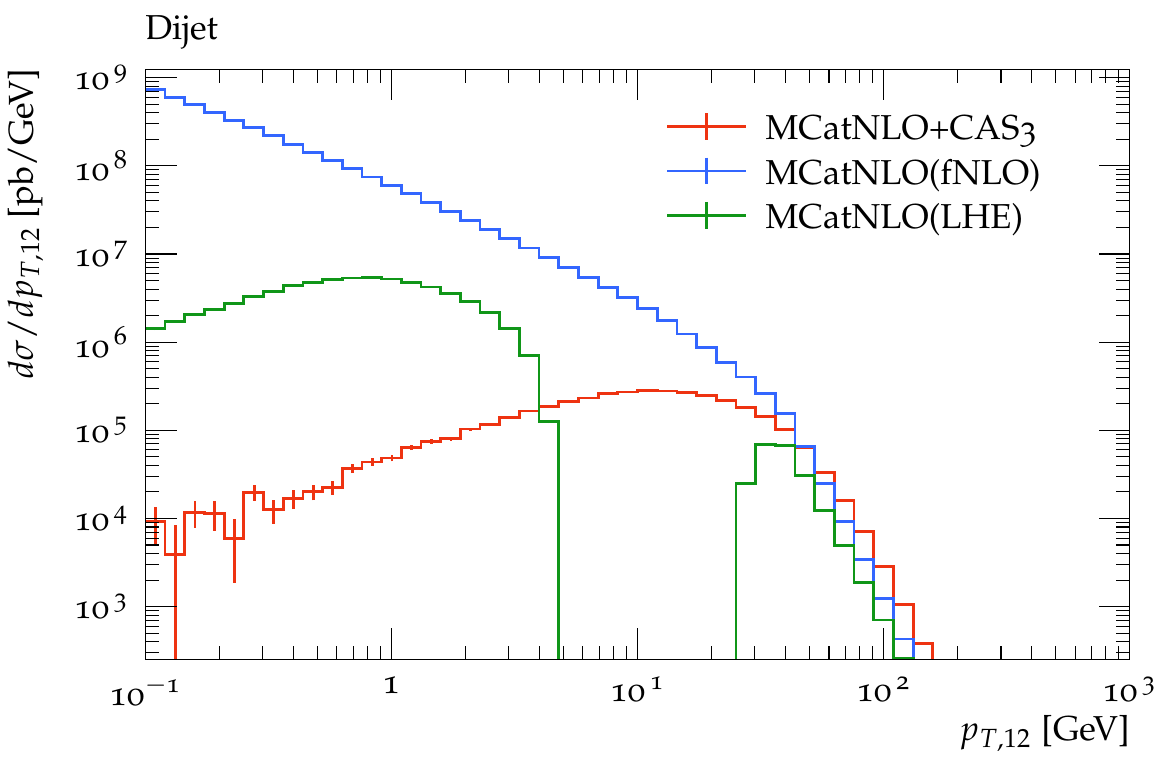} 
\includegraphics[width=0.45\textwidth]{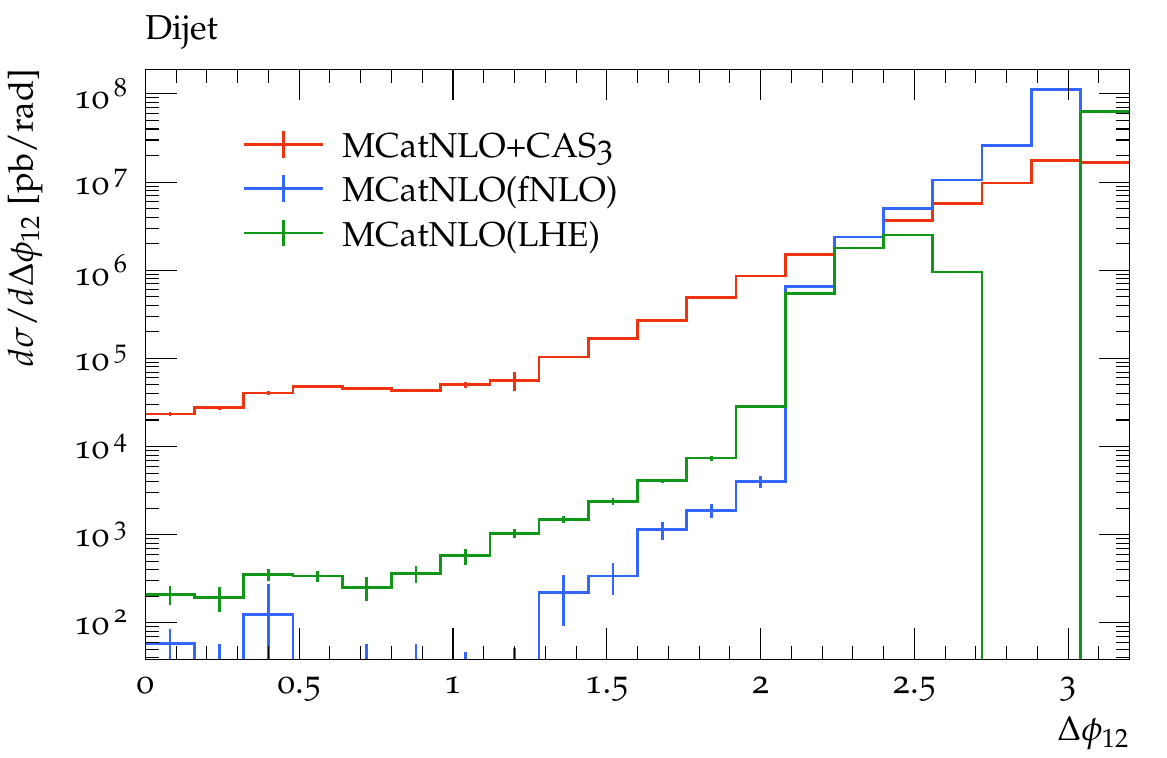} 
  \caption{\small Transverse momentum spectrum of the dijet system \ptjj\ (left)  and $\dphi$ distribution (right).
The predictions are shown for fixed NLO (\mcatnlo (fNLO), the (unphysical) LHE level (\mcatnlo (LHE)) and after inclusion of \PBM -TMDs (\mcatnlo+CAS3). }
\label{mcatnlo_lhe}
\end{center}
\end{figure}

In Fig.~\ref{mcatnlo_lhe} we show results for the   transverse momentum distribution of the dijet system  \ptjj\ and the azimuthal correlation \dphi\ between the two leading jets as obtained from the  \mcatnlo\   calculation at fixed NLO (blue curve), at the level including subtraction terms (LHE level, green curve) and after inclusion of  \PBM -TMDs (red curve).  One can clearly observe the rising cross section of the fixed NLO calculation towards small \ptjj\ (or at large \dphi ).  This is the region in  \ptjj\ and \dphi\    where the subtraction terms are relevant and a physical prediction is obtained when  \PBM -TMDs and parton showers are included. The jets are defined with the anti-\kt\ jet-algorithm~\cite{Cacciari:2008gp}, as implemented in the FASTJET package~\cite{Cacciari:2011ma}, with a distance parameter of R=0.4 and a transverse momentum $\pt > 200 $ \GeV. The use of jets (instead of partons) is the reason for the tail towards small \dphi\  in the \mcatnlo (LHE) and \mcatnlo (fNLO) calculation.

\section{Azimuthal correlations in multijet production}
\label{sec:correlations}

We next apply the framework described in the previous section, based on the matching of   \PBM -TMDs  with  NLO matrix elements, to describe the measurement of azimuthal correlations \dphi\ obtained by CMS at $\sqrt{s}=13$ \TeV~\cite{Sirunyan:2017jnl} and in the back-to-back region ($\dphi \to \pi$)~\cite{Sirunyan:2019rpc}. Only leading jets with a transverse momentum of $\ptmax > 200 $ \GeV\ are considered. We show distributions of \dphi\ for $\ptmax > 200$ \GeV\ as well as for the very high \pt\ region of $\ptmax > 1000$ \GeV , where the jets appear very collimated. We apply the collinear and TMD set PB-NLO-2018-Set~2, unless explicitly specified, with running coupling $\asmz\ = 0.118$. 
We may estimate the  theoretical uncertainties on the predictions by considering two kinds of uncertainties:  those that come from variation of the arbitrary scales that appear in the various factors that enter the jet cross section, and those that come from the determination of the TMD parton distributions and showers. The former include the renormalization scale in the strong coupling, the factorization scale used in the parton distribution and the matching scale to combine the matrix element and PB TMD. The latter include both experimental and model uncertainties in the TMD extraction.  
As regards the scale variations, we present results corresponding to the 7-point scheme variation around the central values for the renormalization and factorization scale (avoiding the extreme cases of variation). We have studied the dependence on the matching scale $\mu_{m}$ and found that is within the band of variation of factorization and renormalization scales.
The experimental and model uncertainties on the determination of the TMD distributions as described in \cite{Martinez:2018jxt} are included. 
\begin{figure}[h!tb]
\begin{center} 
\includegraphics[width=0.45\textwidth]{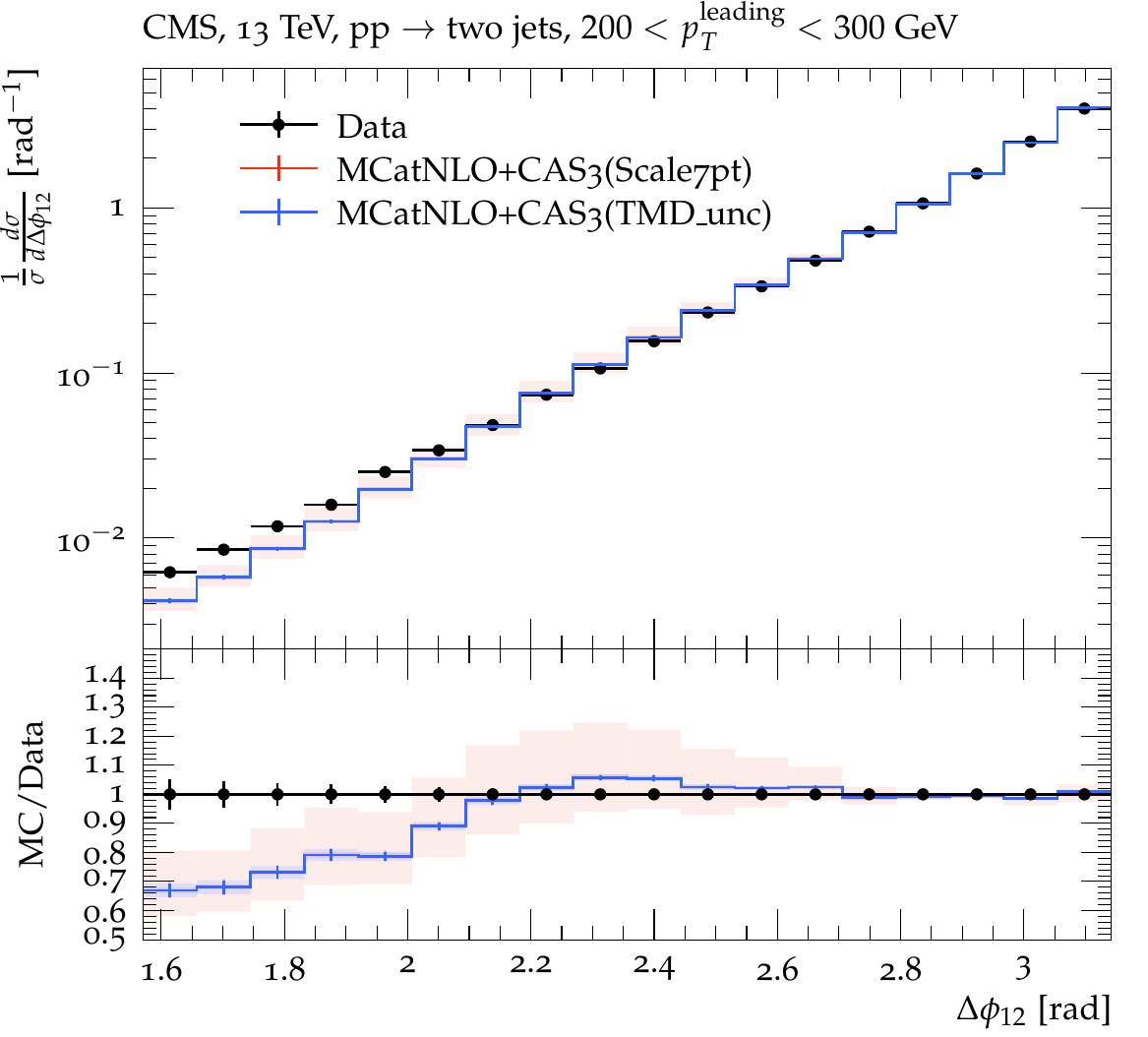} 
\includegraphics[width=0.45\textwidth]{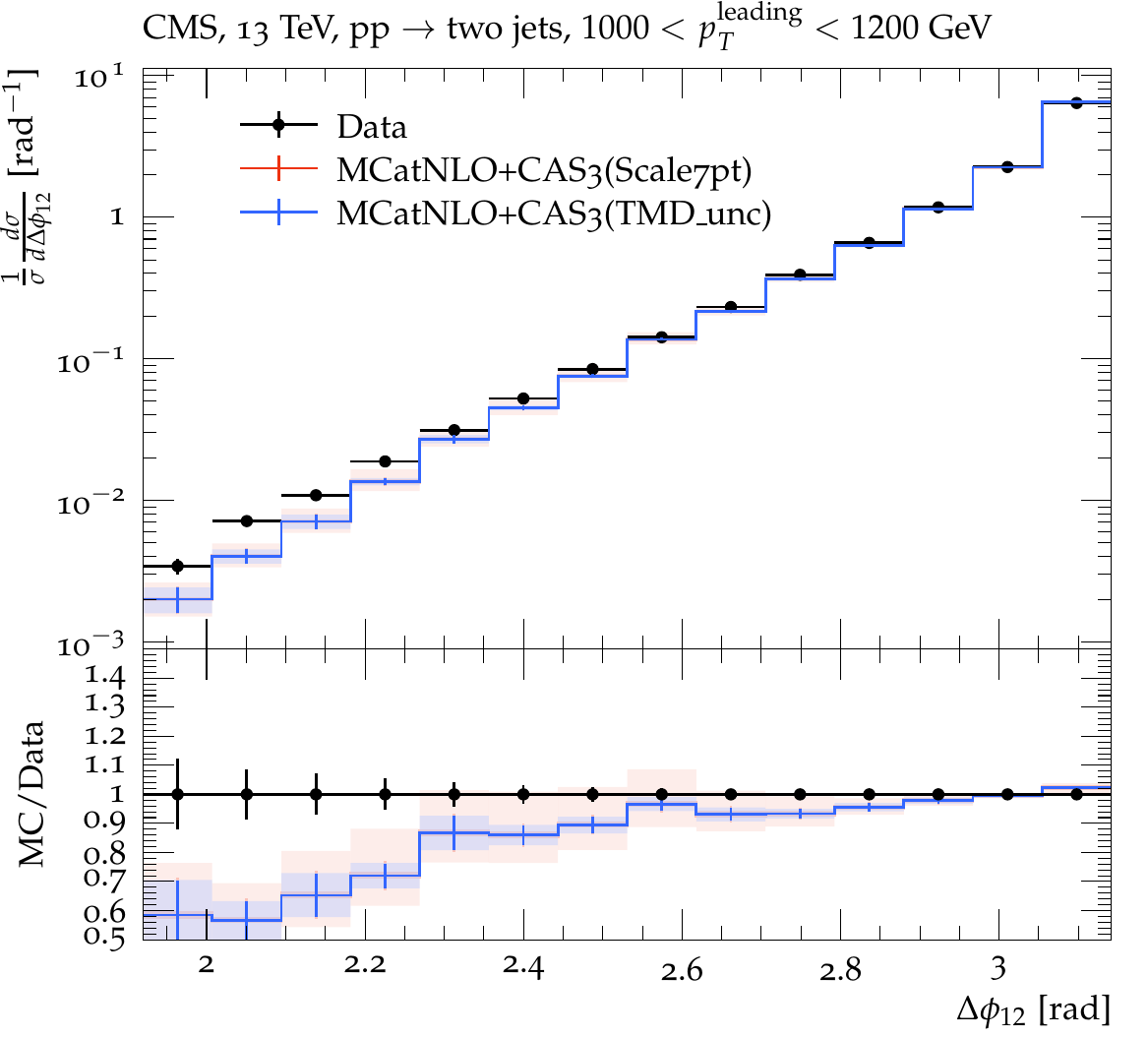} 
  \caption{\small Azimuthal correlation \dphi\ for $\ptmax > 200 $ \GeV\ (left) and $\ptmax > 1000 $ \GeV\ (right) as measured by CMS~\protect\cite{Sirunyan:2017jnl} compared with predictions from \protect\mcatnlo+CAS3. Shown are the uncertainties coming from the scale variation (as described in the text) as well as the uncertainties coming from the TMD.}
\label{dijets_CAS}
\end{center}
\end{figure} 
In Fig.~\ref{dijets_CAS} we show a comparison of the measurement by CMS~\cite{Sirunyan:2017jnl} for different values of \ptmax with the calculation \mcatnlo+CAS3 including \PBM -TMDs, parton shower, and hadronization. The uncertainties from scale variation and TMD determination are shown separately.
\begin{figure}[h!tb]
\begin{center} 
\includegraphics[width=0.45\textwidth]{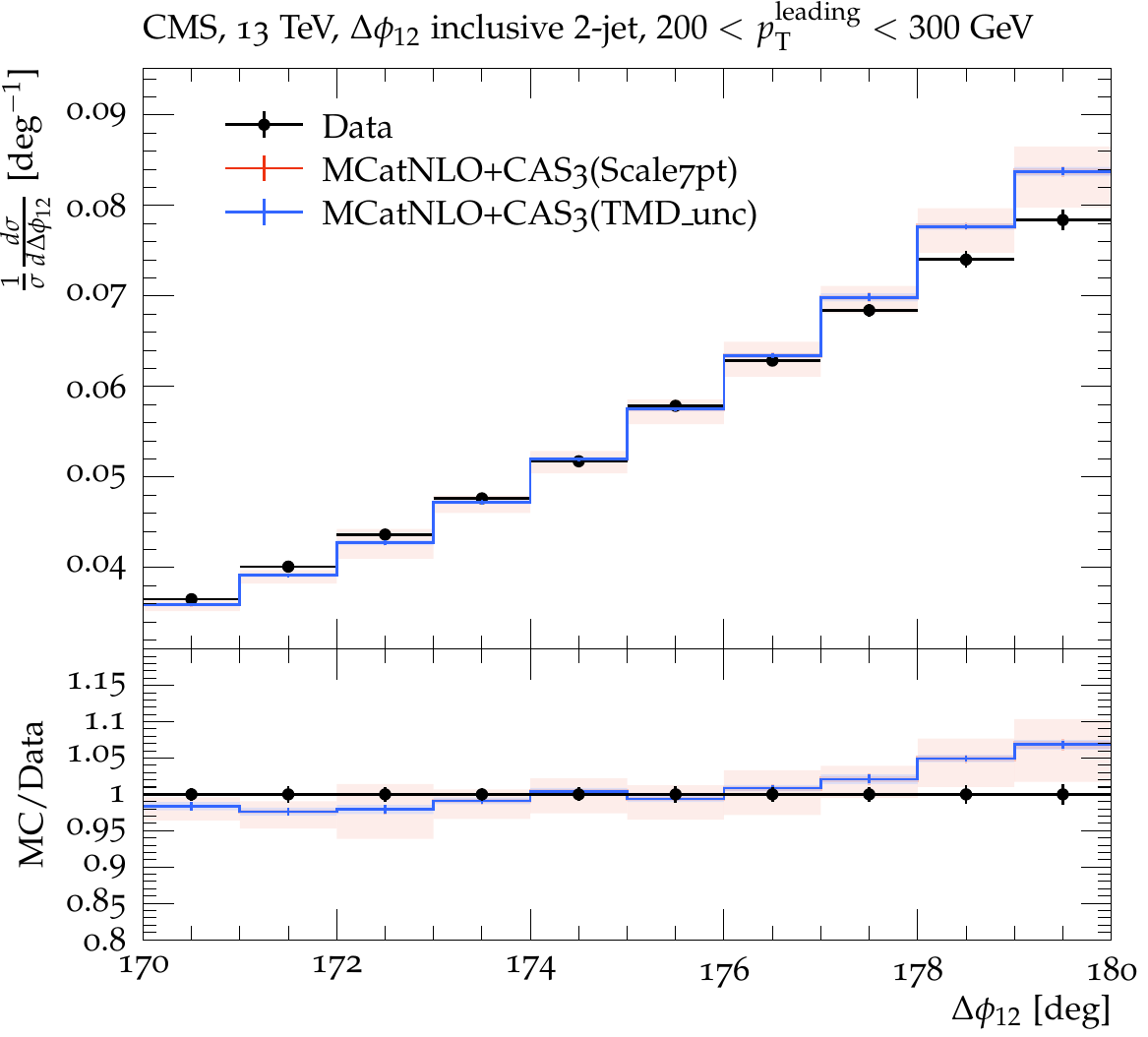} 
\includegraphics[width=0.45\textwidth]{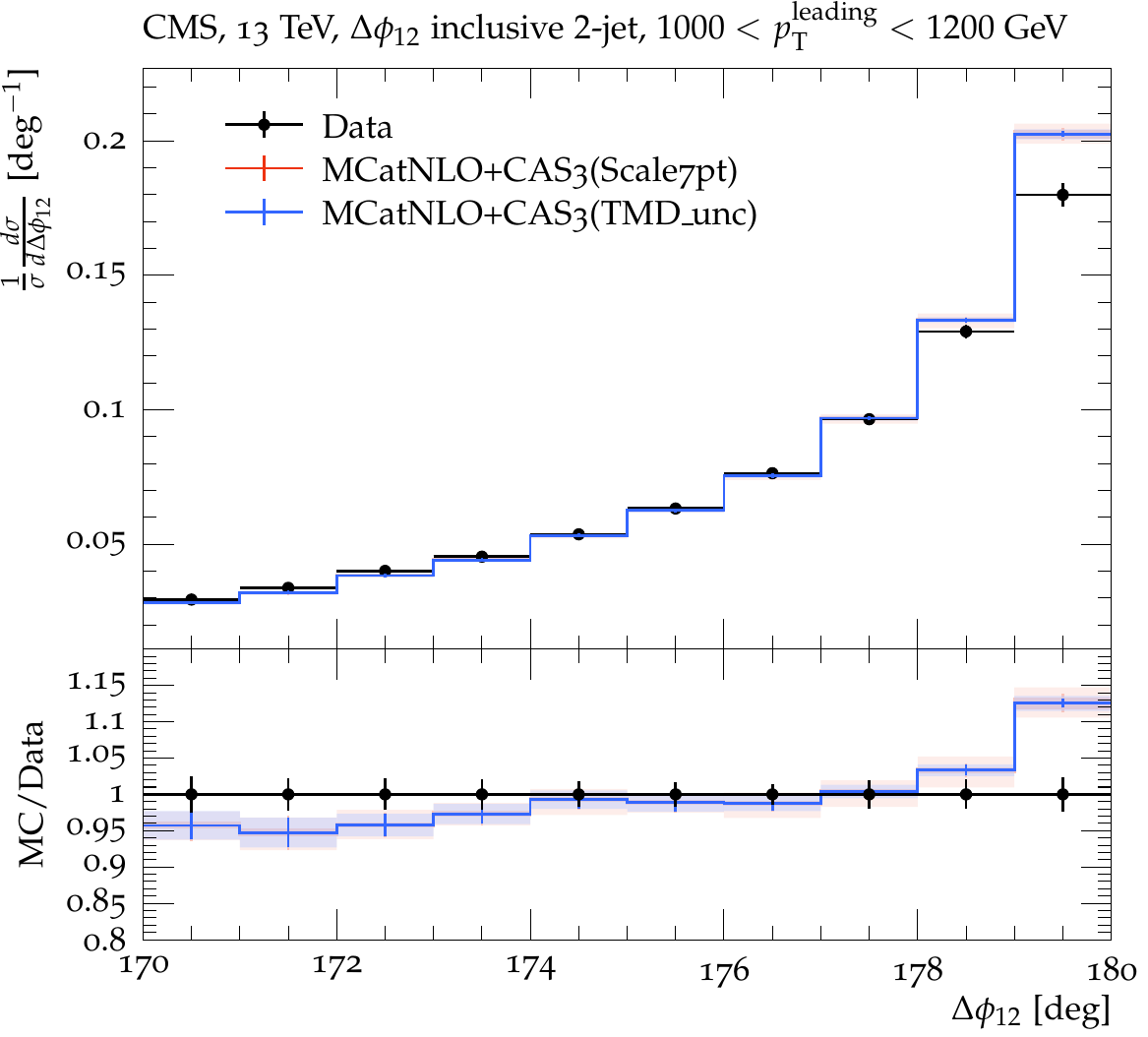} 
  \caption{\small Azimuthal correlation \dphi\  in the back-to-back region for $\ptmax > 200 $ \GeV\ (left) and $\ptmax > 1000 $ \GeV\ (right) as measured by CMS~ \protect\cite{Sirunyan:2019rpc} compared with predictions from \protect\mcatnlo+CAS3. Shown are the uncertainties coming from the scale variation (as described in the text) as well as the uncertainties coming from the TMD.}
\label{b2b-dijets_CAS}
\end{center}
\end{figure} 
In Fig.~\ref{b2b-dijets_CAS}  the measured \dphi\ distribution \cite{Sirunyan:2019rpc} in the back-to-back region is compared with the prediction  \mcatnlo+CAS3.

In general, the measurements are very well described, especially in the back-to-back region. The scale uncertainty is significantly larger than the TMD uncertainty, especially in the low \ptmax region. A difference between the measurement and the prediction is observed for smaller \dphi\ which is due to missing higher order corrections in the matrix element calculation. 
Even at high $\ptmax > 1000 $ \GeV\ the prediction is in agreement with the measurements (within uncertainties), while only in the highest  \dphi\ bin ($\dphi > 179 ^o$) a deviation of about 10\% is observed. 

\begin{figure}[h!tb]
\begin{center} 
\includegraphics[width=0.45\textwidth]{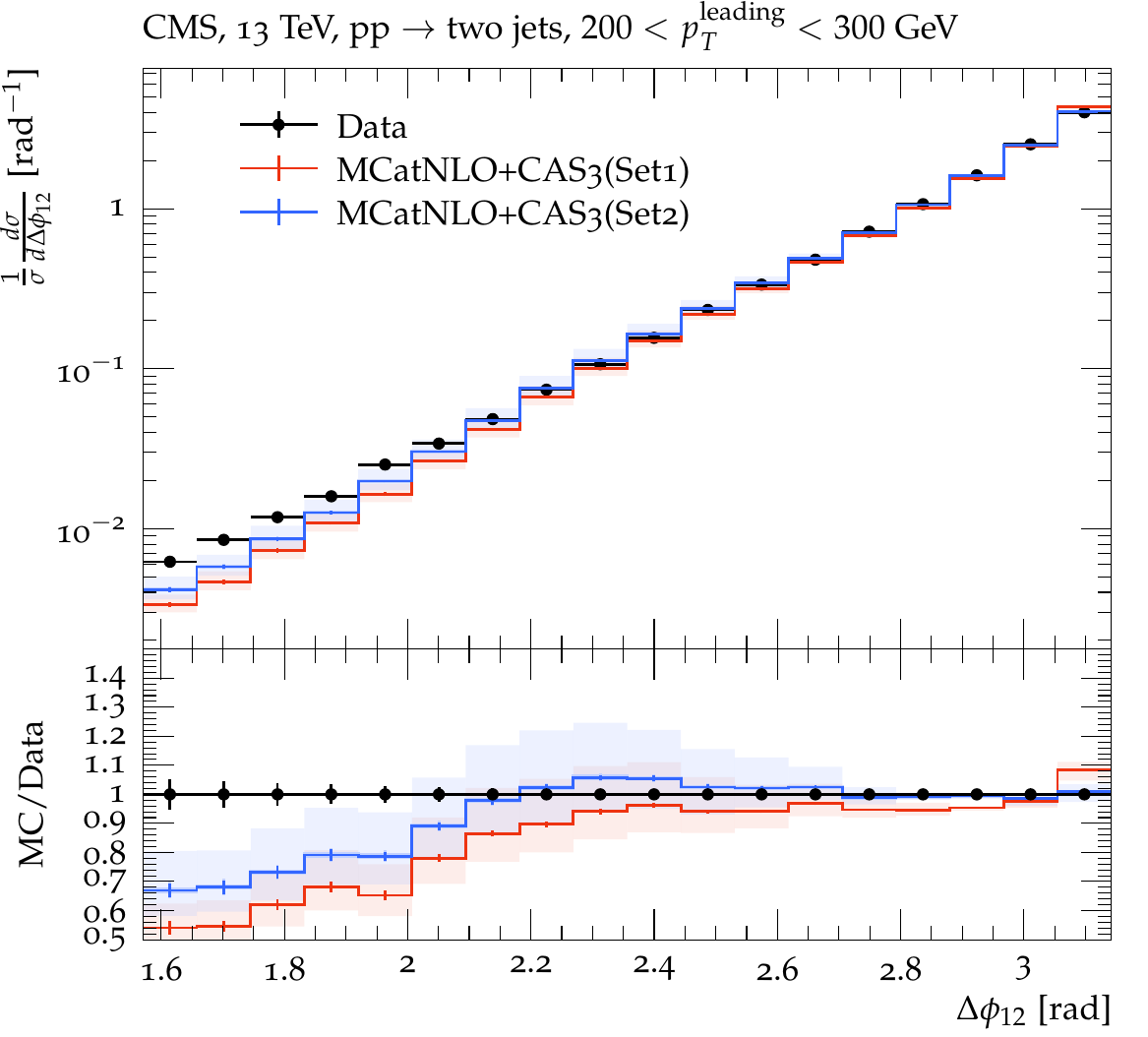} 
\includegraphics[width=0.45\textwidth]{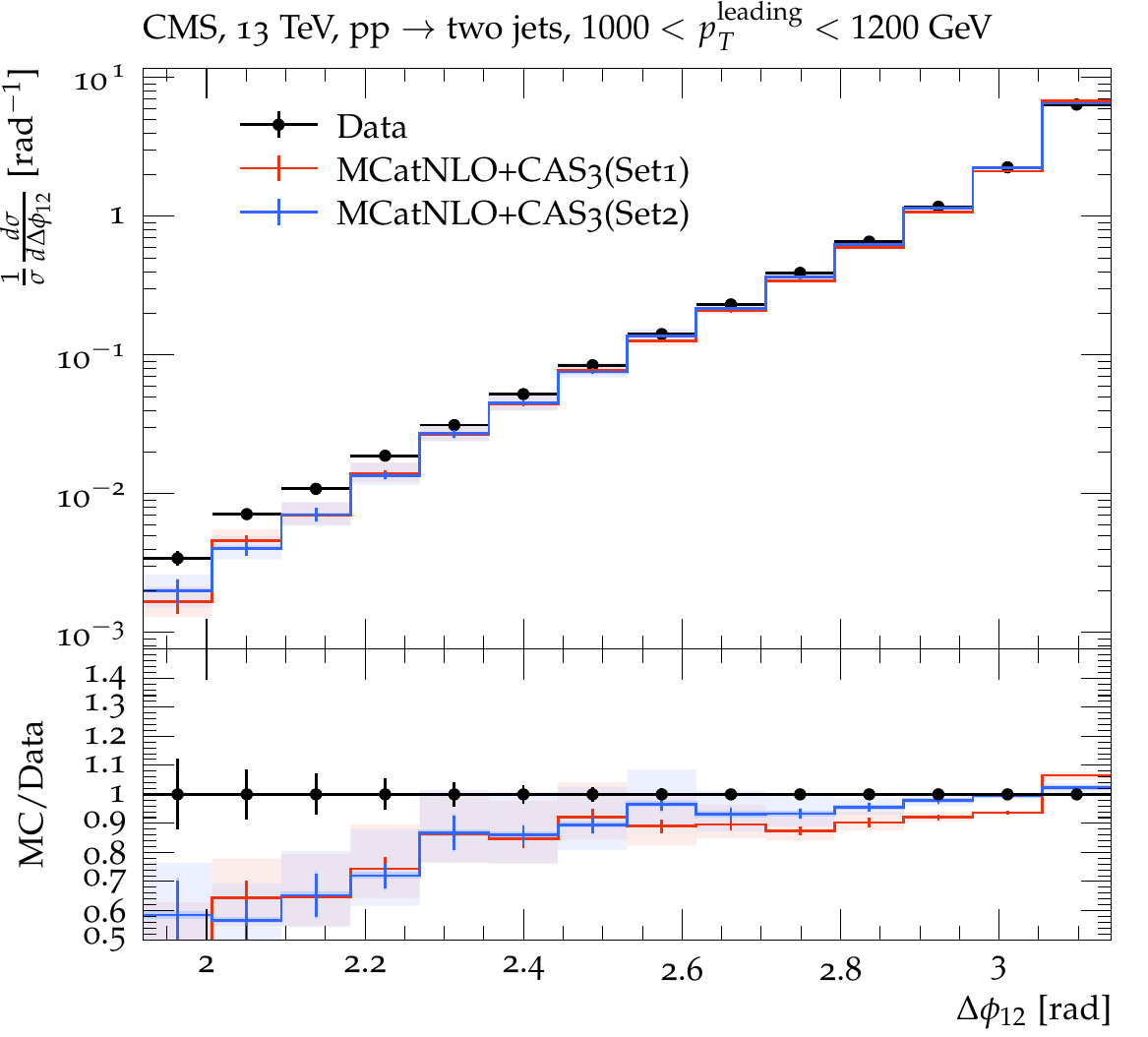} 
  \caption{\small Azimuthal correlation \dphi\ for $\ptmax > 200 $ \GeV\ (left) and $\ptmax > 1000 $ \GeV\ (right) as measured by CMS~\protect\cite{Sirunyan:2017jnl} compared with predictions from \protect\mcatnlo+CAS3. Shown are the uncertainties coming from the scale variation (as described in the text) as well as the uncertainties coming from the TMD.}
\label{dijets_CAS_set12}
\end{center}
\end{figure}

\begin{figure}[h!tb]
\begin{center} 
\includegraphics[width=0.45\textwidth]{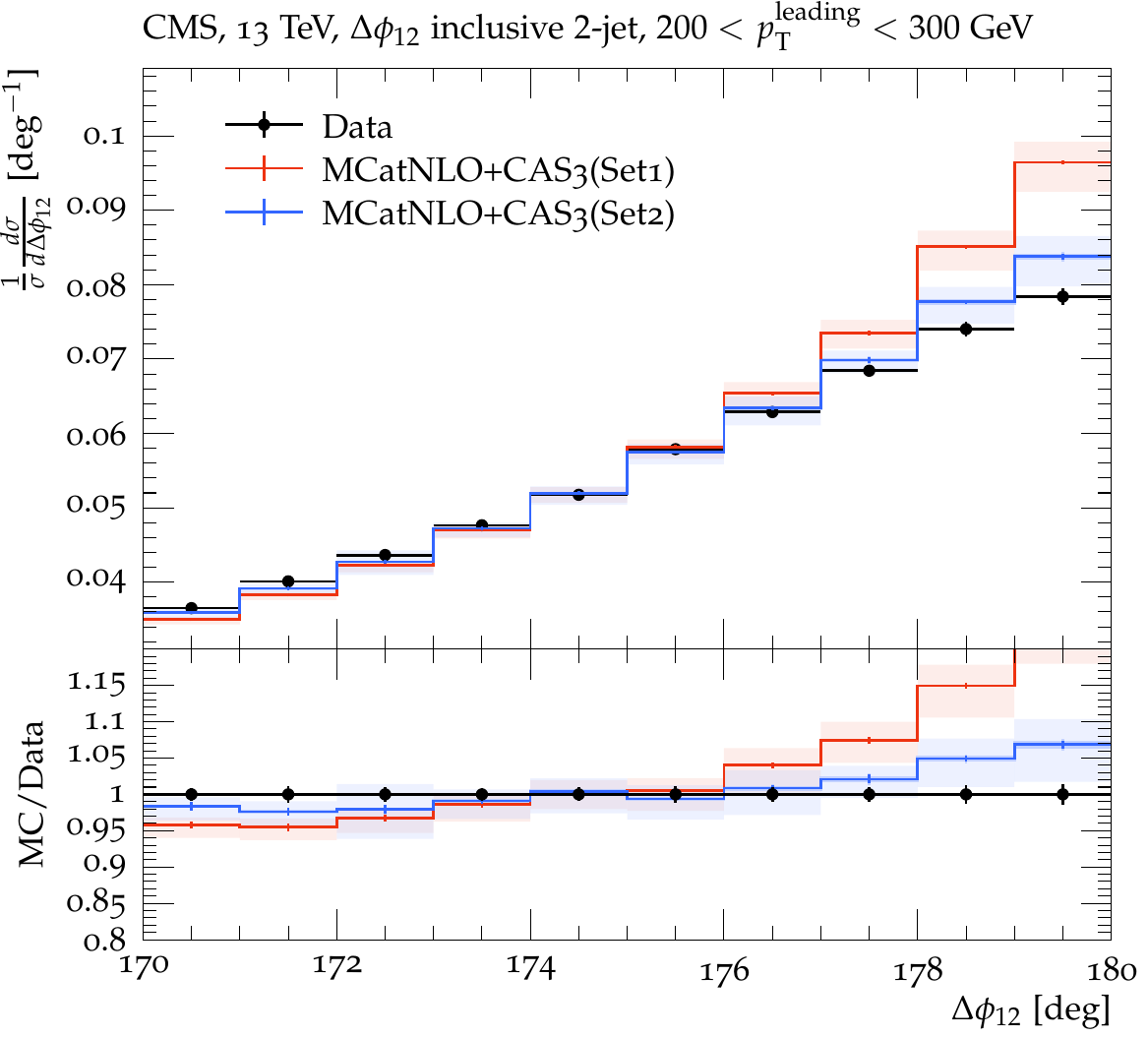} 
\includegraphics[width=0.45\textwidth]{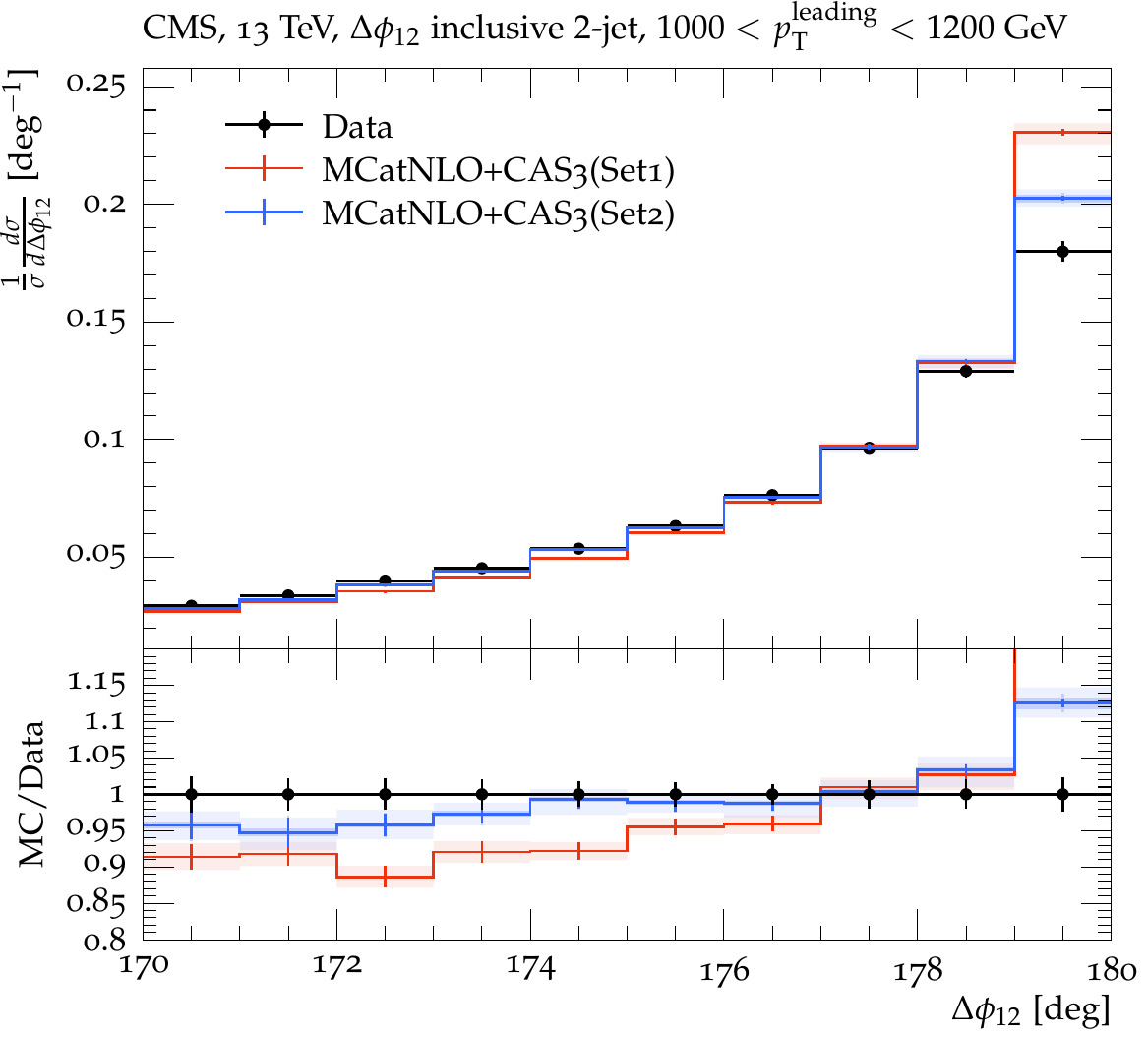} 
  \caption{\small Azimuthal correlation \dphi\  in the back-to-back region for $\ptmax > 200 $ \GeV\ (left) and $\ptmax > 1000 $ \GeV\ (right) as measured by CMS~ \protect\cite{Sirunyan:2019rpc} compared with predictions from \protect\mcatnlo+CAS3. Shown are the uncertainties coming from the scale variation (as described in the text) as well as the uncertainties coming from the TMD.}
\label{b2b-dijets_CAS_set12}
\end{center}
\end{figure} 
In Fig.~\ref{dijets_CAS_set12} and \ref{b2b-dijets_CAS_set12}, the predictions using PB-NLO-2018-Set~1 are compared with those from PB-NLO-2018-Set~2 and with the measurements.
The difference between Set~1 and Set~2 becomes significant in the back-to-back region, which is sensitive to the low \kt -region of the TMD. As already observed in the case of \PZ -boson production in Ref.~\cite{Martinez:2019mwt}, Set~2 with the transverse momentum as a scale for \alphas , which is required from angular ordering conditions, allows a much better description of the measurement. It has been explicitly checked that the choice of the collinear parton density function (in contrast to the choice of the TMD densities) does not matter for the \dphi\ distributions, since they are normalized.
The region of low \dphi\ in Figs.~\ref{dijets_CAS} and~\ref{dijets_CAS_set12} is not well described with an NLO dijet matrix element calculation supplemented with TMD densities and TMD parton shower because in the low  \dphi\ region higher-order hard emissions play a significant role. It has been shown in \cite{Armando-ref2021} that the inclusion of higher order matrix elements with the new TMD merging method of Ref.~\cite{Martinez:2021chk} leads to a very good description of the  low  \dphi\ region.

\begin{figure}[h!tb]
\begin{center} 
\includegraphics[width=0.45\textwidth]{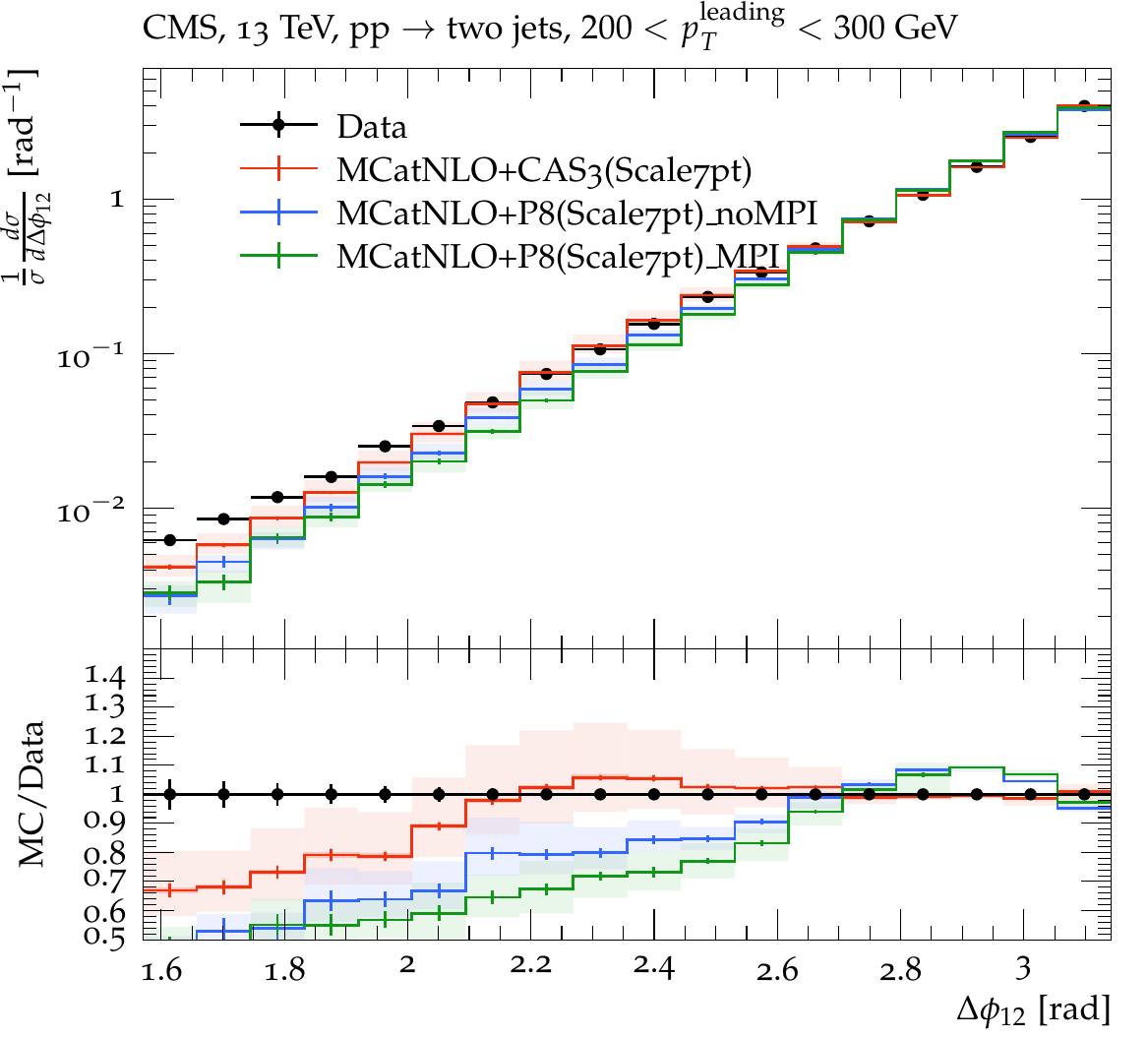} 
\includegraphics[width=0.45\textwidth]{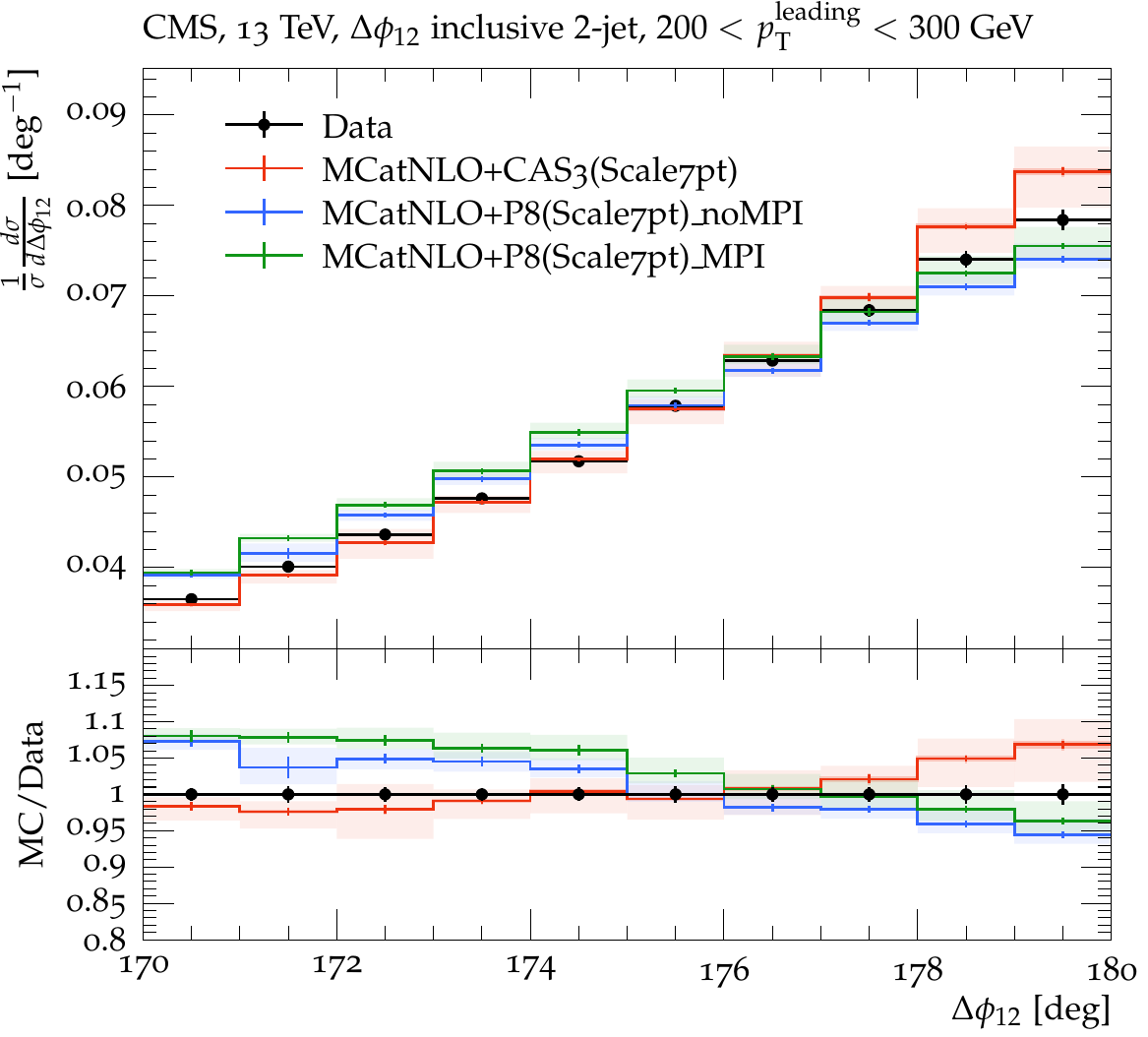} 
  \caption{\small Azimuthal correlation \dphi\  over a wide range and (left)  in the back-to-back region (right) for $\ptmax > 200 $ \GeV\  compared with predictions from \protect\mcatnlo+\pythia 8 and  \protect\mcatnlo+CAS3.
  The uncertainties in the  \protect\mcatnlo+\pythia 8 calculation are obtained from scale and associated shower variations, as described in the main text. }
\label{dijets_CAS_P8}
\end{center}
\end{figure}

\begin{tolerant}{8000}
In Fig.~\ref{dijets_CAS_P8} predictions obtained with \mcatnlo+\pythia 8 are compared with \mcatnlo+CAS3. In the calculation of \mcatnlo+\pythia 8, the  \pythia 8 subtraction terms are used and the NNPDF3.0~\cite{Ball:2014uwa} parton density and tune CUETP8M1~\cite{Khachatryan:2015pea} are applied.
The uncertainties of the PYTHIA prediction are derived by combining the fixed-order scale variation from \mcatnlo\ with renormalization scale variations in the parton shower. We use the method of~\cite{Mrenna:2016sih} together with the guidelines of~\cite{Gellersen:2020tdj} to obtain consistent scale variations where possible. In particular, this means that the renormalization scale variation at fixed order and in the parton shower are fully correlated\footnote{This also ensures that for fixed-order-dominated observables, the cancellation between the expansion of the shower and the subtraction in MC@NLO also occurs for non-central renormalization scales without significant deformation of the -- there fully appropriate -- fixed-order uncertainties.}. The factorization scale variation is only applied at fixed order, as argued in~\cite{Gellersen:2020tdj}.
We observe a significant dependence on the matching scale $\mu_{m}$, the details of matching in case of dijets needs further investigation.

Shown in Fig.~\ref{dijets_CAS_P8} is also the contribution from multiparton interactions, which is very small for jets with $\ptmax > 200 $ \GeV .
The prediction obtained with \mcatnlo+\pythia 8 is in all \dphi\ regions different from the measurement and  \mcatnlo+CAS3, illustrating the role of the treatment of parton showers.
\end{tolerant}

In conclusion, the predictions of  \mcatnlo+CAS3 are in reasonable agreement with the measurements in the larger \dphi\ regions, where the contribution from higher order matrix elements is small. In the back-to-back region ($\dphi \to \pi$) the predictions obtained with \PBM -TMDs and parton shower are in good agreement with the measurement.
The uncertainties of the predictions are dominated by the scale uncertainties of the matrix element calculations, while the \PBM -TMD and TMD shower uncertainties are very small, as they are directly coming from the uncertainties of the \PBM -TMDs. No uncertainties, in addition to those from the \PBM -TMD, come from the \PBM -TMD parton shower.

\section{Conclusion}
\label{sec:concl} 

We have investigated the azimuthal correlation of high transverse momentum jets in $\Pp\Pp$  collisions at $\sqrt{s}=13$ \TeV\ by applying  \PBM -TMD distributions to NLO calculations via \mcatnlo . We use the same \PBM -TMDs and \mcatnlo\ calculations as we have used  for \PZ -production at LHC energies in Ref.~\cite{Martinez:2019mwt}. 
A very good description of the cross section as a function of \dphi\ is observed. In the back-to-back region of $\dphi \to \pi$ a very good agreement is observed with  \PBM -TMD Set~2 distributions~\cite{Martinez:2018jxt} while significant differences are obtained with  \PBM -TMD Set~1 distributions, which use the evolution scale as an argument in \alphas .
This observation confirms the importance of consistently handling the soft-gluon coupling in angular ordered parton evolution.

The uncertainties of the predictions are dominated by the scale uncertainties of the matrix element, while uncertainties coming from the \PBM -TMDs and the corresponding \PBM -TMD shower are very small. No other uncertainties, in addition to those of the \PBM -TMD, come from the \PBM -TMD shower, since it is directly correlated with the \PBM -TMD density.

We have also investigated predictions using \mcatnlo\ with \pythia 8 to illustrate the importance of details of the parton shower.

\vskip 1 cm 
\begin{tolerant}{8000}
\noindent 
{\bf Acknowledgments.} 

\noindent 
This is based in part on studies during the "Virtual Monte Carlo school - PB TMDs with \cascade3"~\cite{MCschool2021_CASCADE} held from 8.-12. November 2021 at DESY, Hamburg.
\noindent 
We are grateful to Olivier Mattelaer from the \MCatNLO\ team for discussions, help and support with the lhe option for fixed NLO calculations in \mcatnlo .
STM thanks the Humboldt Foundation for the Georg Forster research fellowship  and 
gratefully acknowledges support from IPM.
D.~E.~Martins is supported by CNPQ-Brazil, process 164609/2020-2.
\end{tolerant} 
\vskip 0.6cm 

\bibliographystyle{mybibstyle-new.bst}
\raggedright  
\bibliography{/Users/jung/Bib/hannes-bib}

\end{document}